\documentclass[aps,showpacs,nofootinbib,superscriptaddress]{revtex4}

\usepackage{graphicx}
\usepackage{bm}
\usepackage{amsmath}
\usepackage{amssymb}
\usepackage{hyperref}
\allowdisplaybreaks[4]

\newcommand{\g}{\gamma}

\newcommand{\qslash}{\kern 0.2 em q\kern -0.50em /}
\newcommand{\nslash}{\kern 0.2 em n\kern -0.50em /}
\newcommand{\kslash}{\kern 0.2 em k\kern -0.45em /}
\newcommand{\lslash}{\kern 0.2 em l\kern -0.50em /}
\newcommand{\pslash}{\kern 0.2 em p\kern -0.50em /}
\newcommand{\Sslash}{\kern 0.2 em S\kern -0.50em /}
\newcommand{\Pslash}{\kern 0.2 em P\kern -0.50em /}
\newcommand{\Dslash}{\kern 0.2 em D\kern -0.65em /\kern 0.15em}
\newcommand{\slim}{\mskip 1.5mu}

\newcommand{\eps}{\epsilon}

\newcommand{\Tr}{\operatorname*{Tr}\nolimits}

\newcommand{\ii}{i}

\usepackage{color}

\begin{document}

\title{Polarized $\Lambda$ hyperon production in Semi-inclusive deep inelastic scattering off an unpolarized nucleon target}
\author{Yongliang Yang}\affiliation{School of Physics, Southeast University, Nanjing
211189, China}
\author{Zhun Lu}\email{zhunlu@seu.edu.cn}\affiliation{School of Physics, Southeast University, Nanjing
211189, China}
\begin{abstract}
We study the production of polarized $\Lambda$ hyperon in semi-inclusive deep inelastic scattering off an unpolarized target. We include the cases in which the $\Lambda$ hyperon is longitudinally polarized or transversely polarized, and in which the lepton beam is unpolarized or longitudinally polarized. Within the framework of the transverse momentum dependent factorization, we take into account the complete decomposition of the parton correlator for fragmentation up to twist-3. We present the cross section of the process to order $1/Q$. The expressions of the polarized structure functions, which may give rise to various spin asymmetries, are also given.
\end{abstract}

\pacs{13.60.Hb,13.87.Fh,13.88.+e}
\maketitle

\section{Introduction}
Understanding the spin structure of the $\Lambda$ hyperon is one of the most challenging problems in spin physics. The study of the $\Lambda$ polarization dates from 1970s when Fermilab~\cite{Bunce:1976yb} conducted the pioneer measurement in hadronic collision at $300$ GeV.
The measurement of polarization phenomena of $\Lambda$ hyperon production in semi-inclusive deep inelastic scattering (SIDIS) can provide further information about the spin structure of the $\Lambda$ hyperon and the spin-dependent dynamics of fragmentation region, such as the mechanism of spin transfer from outgoing struck quark to a $\Lambda$ hyperon~\cite{Burkardt:1993zh,Jaffe:1996wp,Ma:2000uv,Anselmino:2001ps,Zhou:2009mx}. The polarization of the $\Lambda$ hyperon can be measured by looking at the angle distribution of decay $\Lambda\,\rightarrow \,p\pi$, since the decayed products (proton and pion) will preserve the polarization information of $\Lambda$ hyperon.

The longitudinal polarization of the $\Lambda$ hyperon can be generated in SIDIS by a longitudinally polarized beam off an unpolarized target.
The angular momentum conservation indicates that outgoing quark has the same spin orientation as lepton beam and the polarized quark could fragment into a $\Lambda$ hyperon and transfer its polarization in process. The longitudinal spin transfer has been measured by the HERMES collaboration~
\cite{Airapetian:1999sh,Belostotsky:1999kj,
Bernreuther:2000wf,Airapetian:2006ee,Naryshkin:2007zza,Belostotski:2011zzb}
and the COMPASS collaboration~\cite{Sapozhnikov:2005sc,Alexakhin:2005dz,Alekseev:2009ab}.
The production of transversely polarized $\Lambda$ hyperon in lepton-nucleon scattering has also been proposed as a useful tool to study its spin structure\cite{Anselmino:2001js,Boer:2010ya,Kanazawa:2015jxa}.
Although the transversely polarized $\Lambda$ hyperon production has been measured in hadron collisions~\cite{Bellwied:2002rg,Adler:2002pb,Abelev:2006cs,ATLAS:2014ona}
with different beams, very little experimental information about $\Lambda$ polarization is available from leptoproduction~\cite{Rith:2007zz,Ferrero:2007zz,Airapetian:2014tyc,Karyan:2016rls}.
In SIDIS, polarized $\Lambda$ production is related to quark polarization inside the nucleons as well as the hadronization process in final state.

In this work, we study the semi-inclusive leptoproduction of longitudinally or transversely polarized $\Lambda$ hyperon: $\ell+N\longrightarrow \ell'+\,\Lambda +\,X$, in which a lepton beam (unpolarized or longitudinal polarized) scatters off an unpolarized nucleon target.
To this end, we consider the decomposition of the quark correlation function to the transverse momentum dependent (TMD) parton distribution functions (PDFs) and fragmentation functions (FFs), up to the subleading order of the $1/Q$ expansion.
Within the TMD factorization framework, we compute the parton-model results of the cross section which is differential to the transverse momentum of the $\Lambda$ hyperon.
Particularly, we pay more attention to the T-odd PDFs and FFs, since they play important role on various azimuthal or spin asymmetries in SIDIS.
As shown in Refs.~\cite{Goeke:2003az,Bacchetta:2004zf,Goeke:2005hb}, the presence of the direction of the Wilson line in the decomposition of the parton correlation function will introduce several twist-3 T-odd functions that has not been considered in previous studies~\cite{Mulders:1995dh,Boer:1997nt}.
We will also include these functions to compute the $\Lambda$-spin dependent structure functions.

The paper is organized as follows.
In Sec.~II, we review the decomposition of the parton correlation functions up to twist-3 level.
We then perform the calculation of the hadronic tensor for $\Lambda$ hyperon production in SIDIS with an unpolarized target.
In Sec.~III we decompose the cross section into a complete set of structure functions of the process and provide the spin-dependent structure functions.
Finally, we summarize our results and give conclusions in Sec.IV.

\section{Formalism of the calculation}

In this section, we adopt the approach in Ref.~\cite{Bacchetta:2006tn} to set up the formalism for the calculation of the $\Lambda$ hyperon production in SIDIS
\begin{align}
\ell(l)+ N(P)\rightarrow \ell'(l')+\Lambda(P_\Lambda) + X\,,
\end{align}
where we use $l$, $l'$, $P$ and $P_\Lambda$ to denote the momenta of the incoming lepton beam, the outgoing lepton, the nucleon target $N$ and the $\Lambda$ hyperon, respectively. The momentum of the exchanged virtual photon is defined as $q=l-l'$ and $Q^2=-q^2$. We also define the masses of the nucleon and the $\Lambda$ hyperon as $M$ and $M_\Lambda$.
To express the cross section, we introduce the invariant variables
\begin{align}\label{variable}
  x= \frac{Q^2}{2P\cdot q}\,,~~~~~~y= \frac{P\cdot q}{P\cdot l}\,,~~~~ z= \frac{P\cdot P_\Lambda}{2P\cdot q}\,,~~~\gamma={2Mx\over Q}\,.
\end{align}

\begin{figure*}
  \centering
  \includegraphics[width=0.4\columnwidth]{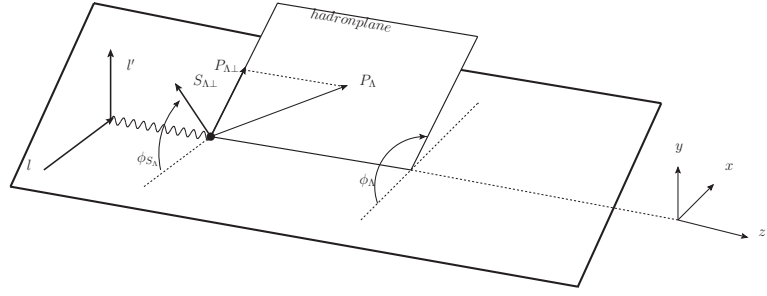}
  \caption{Definition of azimuthal angles for SIDIS in the $\gamma^* \, N$ collinear frame~\cite{Bacchetta:2004jz}, the lepton plane
   is determined by $l$,$l'$. $P_{\Lambda\perp}$ and $S_{\Lambda\perp}$ are transverse parts of $P_\Lambda$ and $S_{h}$ with respect to the photon momentum.}
  \label{sidis}
\end{figure*}

Following the ``Trento convention"~\cite{Bacchetta:2004jz}, we define the azimuthal angles $\phi_h$ of the detected $\Lambda$ hyperon between transverse momentum part and lepton plane as
\begin{align}\label{angles}
  \cos{\phi_\Lambda}=-\frac{l_{\mu}P_{\Lambda\nu}\,g^{\mu\nu}_\perp}{\sqrt{l^2_\perp\,P^2_{\Lambda\perp}}}\,,~~~~~~~~~~
  \sin{\phi_\Lambda}=-\frac{l_{\mu}P_{\Lambda\nu}\,\epsilon^{\mu\nu}_\perp}{\sqrt{l^2_\perp\,P^2_{\Lambda\perp}}},
\end{align}
where $\ell^{\mu}_{\perp} = g^{\mu\nu}_\perp\,\ell_\nu$ and $P^{\mu}_{\Lambda\perp} = g^{\mu\nu}_\perp\,P_{\Lambda\nu}$. We introduce perpendicular projection tensors
\begin{align}\label{tensor}
  g^{\mu\nu}_\perp &\equiv\,g^{\mu\nu}+\hat{z}^\mu\hat{z}^\nu-\hat{t}^\mu\hat{t}^\nu\,,\\
   \eps^{\mu\nu}_\perp &\equiv\,-\eps^{\mu\nu\rho\sigma}\hat{t}_\rho\,\hat{z}_{\sigma}\,,
\end{align}
with nonzero components $g^{11}_\perp = g^{22}_\perp = -1$ and $\eps^{12}_\perp = -\eps^{21}_\perp = 1$. It is convenient to expand the leptonic and hadronic tensor with respect to the virtual photon direction. The two normalized vectors $\hat{t}$ and $\hat{z}$ are
\begin{align}
 \hat{t}^\mu={xP^\mu\over Q}+{q^\mu\over Q}\,,~~~~ \hat{z}^\mu=-{q^\mu\over Q}=-\hat{q}^\mu\,.
\end{align}
We decompose the covariant spin vector $S_\Lambda$ of the $\Lambda$ hyperon as
\begin{align}
 S_\Lambda^\mu=S_{\Lambda\parallel}{P_\Lambda^\mu-q^\mu\,M_\Lambda^2/(P_\Lambda\cdot q)\over M_\Lambda\sqrt{1+Q^2M_\Lambda^2/(P_\Lambda \cdot q)^2}}+S_{\Lambda\perp}^\mu\,.
 \end{align}
The azimuthal angle $\phi_{S_\Lambda}$ relevant for specifying the polarization of $\Lambda$ hyperon is obtained from Eq.~(\ref{angles}) by the replacement: $P_{\Lambda}\rightarrow S_{\Lambda}$. In our study, the target nucleon is unpolarized and the detected $\Lambda$ hyperon is transversely or longitudinally polarized.

The cross section of SIDIS can be expressed as the contraction of the hadronic tensor and the leptonic tensor:
\begin{align}\label{section}
  {{d\sigma}\over{dx dy dz d\phi_\Lambda d\psi dP^2_{\Lambda\perp}}} &=\frac{\alpha^2\,y}{8Q^4\,z}L_{\mu\nu}2MW^{\mu\nu}\,,
\end{align}
where the angle $\psi$ is the azimuthal angle of the outgoing lepton $\ell'$ around beam axis with respect to an arbitrary fixed direction, and we can choose it to be the transverse polarization direction of the $\Lambda$ hyperon.
The expression of the cross section can be simplified with $\phi_{S_\Lambda}$ instead of $\psi$.
In deep inelastic kinematics, one has $d\psi\approx d\phi_{S_{\Lambda}}$~\cite{Boer:1999uu,Diehl:2005pc}.

In the $\gamma^*\,N$ collinear frame~\cite{Mulders:1995dh}, the lepton momentum can be expanded into $\hat{t}$, $\hat{z}$ and the perpendicular components.
Thus, we have the leptonic tensor (neglecting the lepton mass):
\begin{align}\label{lepton}
  L^{\mu\nu}=&{Q^2\over y^2}\bigr[-2(1-y+\frac{1}{2}y^2)g^{\mu\nu}_\perp\,+4(1-y)\hat{t}^\mu\hat{t}^\nu\nonumber\\
  &+4(1-y)(\hat{x}^\mu\hat{x}^\nu+\frac{1}{2}g^{\mu\nu}_\perp)+2(2-y)\sqrt{1-y}\, \hat{t}^{\{\mu}\,\hat{x}^{\nu\}}\nonumber\\
  &-i\lambda\,y(y-2)\eps^{\mu\nu}_\perp-2i\lambda\,y\sqrt{1-y}\, \hat{t}^{[\mu}\eps^{\nu]\rho}_\perp\hat{x}_\rho \bigr]\,,
\end{align}
where $\hat{x}^\mu=l^\mu_\perp/|\bm{l}_\perp|$ is a unit vector in perpendicular direction. The lepton helicity is denoted by $\lambda$. The notations $\{ \}$ and $[\ ]$ indicate symmetrization and
antisymmetrization of Lorentz indices, respectively.

The hadronic tensor in SIDIS is defined as:
\begin{align}\label{hadron}
  2MW^{\mu\nu}=\frac{1}{(2\pi)^3}\sum_X\int{d^3\bm{P}_X\over 2P_X^0}\,\delta^{(4)}(q+P-P_X-P_\Lambda)\langle P|J^\mu(0)|P_{\Lambda}S_\Lambda;P_X\rangle\langle P_{\Lambda}S_\Lambda;P_X|J^\nu(0)|P\rangle\,,
\end{align}
where $J^\mu(\xi)$ is the electromagnetic current. It is understood that the sum $\sum_{X}$ is also over the polarization of undetected hadrons in final state.

\begin{figure*}
  \centering
  \scalebox{0.5}{\includegraphics*[70pt,515pt][1070pt,770pt]{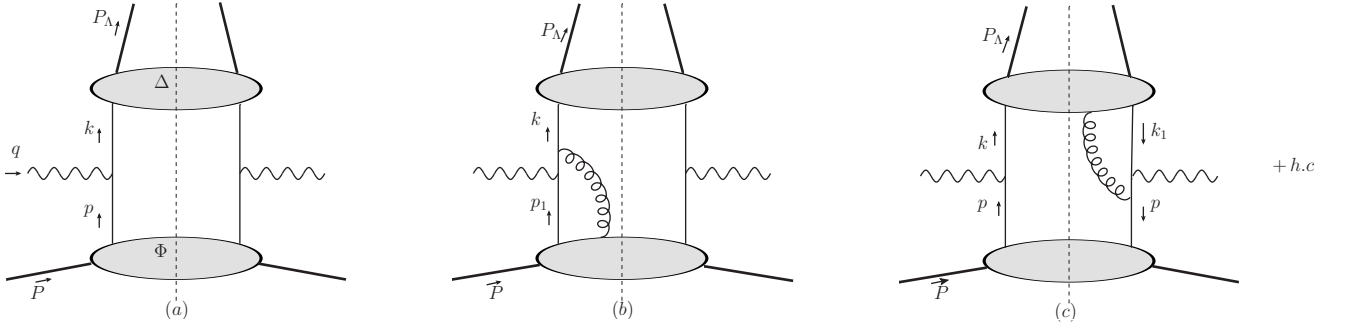}}
  \caption{Diagrams contributing to semi-inclusive DIS up to $\mathcal{O}(1/Q)$, the ``h.c." stands for the hermitian conjugation of (b) and (c).}
  \label{contri}
\end{figure*}

In the factorization framework~\cite{Collins:2004nx,Ji:2004xq,Ji:2004wu,Aybat:2011zv}, the cross section of SIDIS can be written as the convolution of the lepton-quark scattering process (the hard part) and non-perturbative TMDs (the soft part)~\cite{Curci:1980uw,Collins:1997sr,Collins:1989gx,Collins:1985ue,Ma:2008cj}.
At the tree level,
the hadronic tensor can be factorized in terms of various TMD PDFs and FFs up to sub-leading twist in the sense of $1/Q$ expansion. Therefore, the hadronic tensor can be obtained from the diagrams shown in Fig.~\ref{contri}.
Here, Fig.~\ref{contri}a only involves quark-quark matrix elements, and Fig.~\ref{contri}b and Fig.~\ref{contri}c involve quark-gluon-quark matrix elements.
The ``h.c." represents the diagrams hermitian conjugate to Fig.~\ref{contri}b and Fig.~\ref{contri}c, with gluon attaching to the other side of the final state cut.
Up to $\mathcal{O}(1/Q)$, the corresponding contributions of the hadronic tensor can be expressed as~\cite{Mulders:1995dh,Bacchetta:2004zf,Boer:2003cm}
\begin{align}
&2M W^{\mu\nu}=\sum_a e^2_a\int{d^2\bm p_T\,d^2\bm k_T}\,\delta^2(\bm p_T-\bm k_T+\bm q_T)\,\Tr{\bigg\{}\Phi^a(x,p_T)\gamma^\mu\Delta^a(z,k_T)\gamma^\nu\,\nonumber\\
&-\gamma^\alpha {\nslash_+\over Q\sqrt{2}} \gamma^\nu \tilde{\Phi}^a_{A\alpha}(x,p_T)\gamma^\mu\Delta^a(z,k_T)\,-\gamma^\alpha {\nslash_-\over Q\sqrt{2}} \gamma^\mu \tilde{\Delta}^a_{A\alpha}(z,k_T)\gamma^\nu\Phi^a(x,p_T)\,-\,h.c.{\bigg\}}\,.
\label{eq:hadtensor}
\end{align}
In Eq.~(\ref{eq:hadtensor}), the terms with $n_+$ and $n_-$ arise from fermion propagators in the quark-lepton scattering part with corrections of order $1/Q$.~\cite{Boer:1997mf,Boer:2003cm}

Note that here we decompose the spin and momentum vectors using two light-like vectors $n_+$ and $n_-$ in the light-cone coordinate, in which the transverse direction is defined in the $\Lambda\,N$ collinear frame (T-vectors).
Particularly, $P$ and $P_\Lambda$ have no transverse momentum part, and they can be decomposed as
\begin{align}
P^\mu = P^+n^\mu_+\,+{M^2\over 2P^+}n^\mu_-,~~~~~~~~~~~~~~P_\Lambda^\mu = P_\Lambda^-n^\mu_-\,+{M_\Lambda^2\over 2P_\Lambda^-}n^\mu_+,
\end{align}
 with $n_+$ and $n_-$ can be expressed in terms of $\hat{t}$, $\hat{z}$ and the transverse component $q_T$,
 \begin{align}
n^\mu_+ = {1\over\sqrt{2}}(\hat{t}^\mu +\hat{z}^\mu),~~~~~~~~~~~~~~n^\mu_- = {1\over\sqrt{2}}(\hat{t}^\mu-\hat{z}^\mu-2{q^\mu_{T}\over Q})\,.
\end{align}
Thus the relation between the two bases of the transverse vectors can be obtained from the following expression~\cite{Boer:2003cm,Mulders:1995dh}:
\begin{align}
g^{\mu\nu}_T = g^{\mu\nu}_{\perp}-{Q_T\over Q}\hat{q}^{\{\mu}\hat{x}^{\nu\}}+ {Q_T\over Q}\hat{t}^{\{\mu}\hat{x}^{\nu\}}.
\end{align}
The decomposition of the spin vector $S_\Lambda$ has the form
\begin{align}
S_{\Lambda}^\mu = S_{\Lambda\,L}{(P_\Lambda\cdot n_+)n_-^\mu\,- (P_\Lambda\cdot n_-)n_+^\mu\over M_\Lambda}+ S_{\Lambda\,T}^\mu\,,
\end{align}

To construct the hadronic tensor, we start from the general structure of the correlation functions~\cite{Metz:2016swz} shown in Eq.~(\ref{eq:hadtensor}), which are $\Phi$ for the quark distributions, $\Delta$ for the quark fragmentation, and the quark-gluon-quark correlators $\tilde{\Phi}_A$ and $\tilde{\Delta}_A$.

The quark-quark distribution correlation function for unpolarized nucleon in SIDIS is defined as
 \begin{align}
\Phi(x,p_T) =\int\frac{d\xi^- d^2\bm{\xi}_T}{(2\pi)^{3}}\,e^{i p \cdot \xi}\,\langle P|\,\bar{\psi}(0)\,{\cal U}^{n_-}_{(0,+{\infty})} {\cal U}^{n_-}_{(+{\infty},\xi)}\,\psi(\xi)|P\rangle \bigg|_{\xi^+=0}\,.
\label{eq:delta}
  \end{align}
In this correlator, the Wilson lines are given as
\begin{align}
\mathcal{U}^{n_{-}}_{(+\infty,\xi)} \equiv &\mathcal{U}^{T}_{(\infty_T,\xi_T;+\infty^{\pm})}\,\mathcal{U}^{n_{-}}_{(+\infty^{-},\xi^-;\xi_T)},\\
\mathcal{U}^{n_{-}}_{(0,+\infty)} \equiv &\mathcal{U}^{n_{-}}_{(\xi^-,+\infty^{-};0_T)}\,\mathcal{U}^{T}_{(0_T,\infty_T;+\infty^{-})}\,,
\end{align}
where the superscript $n_-$ of $\mathcal{U}$ indicates a Wilson line running along the minus direction in SIDIS.
Detailed definitions of the Wilson lines for TMDs can be found in Refs.~\cite{Belitsky:2002sm,Ji:2002aa}.
Particularly, the definitions of Wilson lines for the correlation functions can differ in different processes~\cite{Collins:2002kn,Bacchetta:2005rm,Bomhof:2004aw,Bomhof:2006dp}.
For instance, all occurrences of $\infty^-$ in Wilson line in SIDIS should be replaced by $-\infty^-$ in Drell-Yan process.

The quark-quark correlator for an unpolarized nucleon can be decomposed as~\cite{Bacchetta:2004zf,Goeke:2005hb}
\begin{align}
 \label{eq:distribution}
\Phi(x, p_T) =&{1\over 2} \biggl\{ f_1 \nslash_+ + \ii h_1^\perp \frac{\bigl[\pslash_T, \nslash_+ \bigr]}{2M}\biggr\}\nonumber \\
 +&\frac{M}{2P^+}\,\biggl\{ e + f^\perp \frac{\pslash_T}{M} - g^\perp\,\g_5\frac{\eps_T^{\rho \sigma} \g_{\rho}\slim p_{T \sigma}}{M} +\ii h\frac{ \bigl[\nslash_+,\nslash_-\bigr]}{2} \biggr\},
\end{align}
where we limit the expression to the leading and subleading terms in $1/Q$ expansion.

The fragmentation correlator $\Delta$ is defined as~\cite{Collins:2011zzd,Boer:2003cm,Pitonyak:2013dsu,Collins:1992kk}
  \begin{align}
\Delta(z,k_T) =\frac{1}{2z}\sum_X\,\int\frac{d\xi^+ d^2\bm{\xi}_T}{(2\pi)^{3}}\;e^{i k \cdot \xi}\,\langle 0|\, {\cal U}^{n_+}_{({\infty},\xi)}\,\psi(\xi)|P_\Lambda,S_\Lambda; X\rangle\langle P_\Lambda,S_\Lambda; X|\bar{\psi}(0)\,{\cal U}^{n_+}_{(0,+{\infty})}|0\rangle \bigg|_{\xi^-=0}\,,
\label{eq:delta}
  \end{align}
with $k^-={P_\Lambda^-\over z}$, where the momentum fraction $z$ in fragmentation functions coincide with the variable defined in Eq.~(\ref{variable}). An similar remark holds for the Bjorken variable $x$ in the definition of distribution functions.

Up to twist-3 level, a complete parameterization of the fragmentation correlator $\Delta$ complying with hermiticity and parity constraints can be given as~\cite{Metz:2016swz}
\begin{align}
 \label{eq:corr}
\Delta(z, k_T) =&{1\over 2}\biggl\{ D_1 \nslash_- + D^\perp_{1T}{\epsilon^{\rho\sigma}_T\,k_{T\rho}S_{\Lambda\,T\sigma}\over M_\Lambda} +G_{1s}\gamma_5\nslash_- \nonumber\\
&+ H_{1T}\frac{\bigl[\Sslash_{\Lambda T}, \nslash_- \bigr]\gamma_5}{2}+ H^\perp_{1s}\frac{\bigl[\kslash_{T}, \nslash_- \bigr]\gamma_5}{2M_\Lambda}+ \ii H_1^\perp \frac{\bigl[\kslash_T, \nslash_- \bigr]}{2M_\Lambda}\biggr\}\nonumber \\
 +&\frac{M_\Lambda}{2P_\Lambda^-}\,\biggl\{ E -\ii E_s\gamma_5\,+E^\perp_{T}\frac{\eps_T^{\rho \sigma} k_{T\rho} S_{\Lambda\,T \sigma}}{M_\Lambda}\nonumber\\
 &+D^\perp \frac{\kslash_T}{M_\Lambda}+D^\prime_T\eps_T^{\rho \sigma} \g_{\rho}S_{\Lambda\,T\sigma}+D^\perp_s\frac{\eps_T^{\rho \sigma} \g_{\rho}\slim k_{T \sigma}}{M_\Lambda}\nonumber\\
 &+G^\prime_T\gamma_5\Sslash_{\Lambda\,T} +G^\perp_s\g_5{\kslash_T\over M_\Lambda} + G^\perp\,\g_5\frac{\eps_T^{\rho \sigma} \g_{\rho}\slim k_{T \sigma}}{M_\Lambda}\nonumber\\
 &+H_s\frac{\bigl[\nslash_-, \nslash_+ \bigr]\g_5}{2}+H^\perp_T\frac{ \bigl[\Sslash_{\Lambda T}, \kslash_T \bigr]\g_5}{2M_\Lambda}+\ii H\frac{ \bigl[\nslash_-, \nslash_+ \bigr]}{2} \biggr\}\,,
\end{align}
where the FFs on the r.h.s. depend on $z$ and $k_T^2 \equiv -\bm k_T^2$.
We use the shorthand notation~\cite{Mulders:1995dh} for the function $G_{1s}$:
 \begin{align}\label{instead}
    G_{1s}=S_{\Lambda\,L}G_{1L}-{k_T\cdot\,S_{\Lambda\,T}\over M_\Lambda}G_{1T}
  \end{align}
and so forth for the other functions.
It is easy to find that the decomposition of $\Delta$ is obtained from that of $\Phi$ by the replacements:
   \begin{align}\label{instead}
   n_+\leftrightarrow n_-,~~\eps_T\rightarrow -\eps_T,~~P^+\rightarrow P_\Lambda^-,~~M\rightarrow M_\Lambda,~~x\rightarrow 1/z\,,
  \end{align}
and the PDFs are replaced by the corresponding FFs (e.g. with $f_1$ replaced by $D_1$ and all other letters are capitalized).

As shown in Eq.~(\ref{eq:corr}), there are three extra twist-3 T-odd FFs, denoted by $E^\perp_T$, $D^\perp_T$ and $G^\perp$, which have not been presented in Refs.~\cite{Mulders:1995dh,Boer:1997mf}.
These functions, analogous to the TMD PDFs $e^\perp_T$, $f^\perp_T$ and $g^\perp$, arise due to the presence of the direction of the Wilson line $n_+$ in the unintegrated correlator (\ref{eq:corr}).
Among them, $D^\perp_T$ is introduced to maintain the symmetry with other functions and simplify the expression of the results~\cite{Goeke:2005hb}.
As we have few experimental information on T-odd fragmentation functions, model calculation is an important way to acquire knowledge of these quantities, such as the spectator models~\cite{Bacchetta:2007wc,Yang:2016mxl,Lu:2015wja}.
In SIDIS, the contributions of the functions $g^\perp$ and $G^\perp$ have to be taken into account, and they could provide useful explanation for the difference of the $A_{LU}$ and $A_{UL}$ asymmetries~\cite{Bacchetta:2004zf}.
Finally, $E^\perp_T$, $D^\perp_T$ are polarized fragmentation functions appearing in polarized $\Lambda$ hyperon production in SIDIS.

At last, we examine the quark-gluon-quark correlator $\tilde{\Phi}_A^\alpha$ and $\tilde{\Delta}_A^\alpha$ appearing in the last line of Eq.~(\ref{eq:hadtensor})
\begin{align}
\begin{split}\label{eq:qgq1}
\tilde{\Phi}_A^\alpha(x,p_T) &=\int \frac{d\xi^{-}d^2\bm\xi_T} {(2\pi)^3}\int  e^{\ii p\cdot \xi} \langle P|\bar{\psi}(0) \int^{\xi^-}_{\pm\infty} d{\eta^-}\mathcal{U}^{n_-}_{(0,+\infty)}gF^{+\alpha}_\perp (\eta) \mathcal{U}^{n_-}_{(\eta,\xi)}\mathcal{U}^{n_-}_{(+\infty,\eta)}\psi(\xi)|P\rangle\bigg|_{\begin{subarray}{l}
\eta^+ = \xi^+=0 \\ \eta_T = \xi_T \end{subarray}}\,,
\end{split} \displaybreak[0]\\
\begin{split}
\tilde{\Delta}_A^\alpha(z,k_T) &=\frac{1}{2z}\sum_{X}\hspace{-0.55cm}\int \;\int \frac{d\xi^{+}d^2\bm\xi_T} {(2\pi)^3}\int  e^{\ii k\cdot \xi} \langle 0| \int^{\xi^+}_{\pm\infty^+} d{\eta^+}\mathcal{U}^{n_+}_{(+\infty,\eta)}\\
 &\times gF^{-\alpha}_\perp (\eta) \mathcal{U}^{n_+}_{(\eta,\xi)} \psi(\xi)|P_\Lambda,S_\Lambda;X\rangle\langle P_\Lambda,S_\Lambda;X|\bar{\psi}(0)\mathcal{U}^{n_+}_{(0,+\infty)}|0\rangle\bigg|_{\begin{subarray}{l}
\eta^- = \xi^-=0 \\ \eta_T = \xi_T \end{subarray}}\,.
\label{eq:qgq2}
\end{split}
\end{align}
Compared to $\Phi$ and $\Delta$, $\tilde{\Phi}^\alpha_A$ and $\tilde{\Delta}_A^\alpha$ contain an additional gluon leg~\cite{Boer:2003cm,Metz:2012ct,Boer:1997mf}.
$\tilde{\Phi}^\alpha_A$ can be decomposed as
\begin{align}
\label{eq:qgqdist}
\tilde{\Phi}^\alpha_A(x, p_T) =&{xM\over 2} \biggl\{ [(\tilde{f}^\perp-i\tilde{g}^\perp){p_{T\rho}\over M}](g^{\alpha\rho}_T-i\epsilon_T^{\alpha\rho}\gamma_5)+(\tilde{h}+i\tilde{e})i\gamma^\alpha_T +\cdots (g^{\alpha\rho}_T+i\epsilon_T^{\alpha\rho}\gamma_5)\biggr\} {\nslash_+\over 2}\,,
\end{align}
where we consider the target nucleon to be unpolarized.

After performing the replacements in Eq.~(\ref{instead}), we can also decompose the quark-gluon-quark correlator $\tilde{\Delta}^\alpha_A$ as follows
\begin{align}
\label{eq:qgqfrag}
\tilde{\Delta}^\alpha_A (z,k_T)=&{M_\Lambda\over 2z}\bigg{\{}\big[(\tilde{D}^\perp-i\tilde{G}^\perp){k_{T\rho} \over M_\Lambda}+(\tilde{D}^\prime_T+i\tilde{G}^\prime_T)\epsilon_{T\rho\sigma}S_{\Lambda\,T}^{\sigma}+(\tilde{D}^\perp_s+\ii\tilde{G}_s^\perp){\epsilon_{T\rho\sigma} k_T^\sigma \over M_\Lambda}\big](g_T^{\alpha\rho}+i\epsilon_T^{\alpha\rho}\gamma^5)\,\nonumber\\
 &-(\tilde{H}_s+i\tilde{E}_s) \gamma^\alpha_T \gamma^5+ [(\tilde{H}+i\tilde{E})-(\tilde{H}_T^\perp-i\tilde{E}_T^\perp){\epsilon_{T}^{\rho\sigma} k_{T\rho} S_{\Lambda\,T\sigma} \over M_\Lambda}]i\gamma^\alpha_T+\cdots\bigg{\}}{\nslash_-\over 2}\,.
\end{align}
It is convenient to parameterize the following combinations as
\begin{align}
\begin{split}
{z\over 2M_\Lambda}\Tr[\tilde{\Delta}_{A\alpha}\,\sigma^{\alpha\,-}] &= \tilde{H}+ i\tilde{E}-{\epsilon_T^{\rho\sigma}k_{T\rho}S_{\Lambda\,T\sigma}\over M_\Lambda}(\tilde{H}^\perp_T-i\tilde{E}^\perp_T),
\end{split} \displaybreak[0]\\
\begin{split}
{z\over 2M_\Lambda}\Tr[\tilde{\Delta}_{A\alpha}\,i\sigma^{\alpha\,-}\gamma_5] &= \tilde{H}_s+ i\tilde{E}_s,
\end{split}\displaybreak[0]\\
\begin{split}
{z\over 2M_\Lambda}\Tr[\tilde{\Delta}_{A\rho}(g^{\alpha\rho}_T-i\epsilon_T^{\alpha\rho}\gamma_5)\gamma^-] &= {k_T^\alpha\over M_\Lambda}(\tilde{D}^\perp-i\tilde{G}^\perp )+{\epsilon_T^{\alpha\rho}S_{\Lambda T\rho}}(\tilde{D}_T+i\tilde{G}_T)\\
&+{\epsilon_T^{\alpha\rho}k_{T\rho}\over M_\Lambda}(\tilde{D}^\perp_s+i\tilde{G}^\perp_s)\,,
\end{split}
\end{align}
where the indices $\alpha$, $\rho$ and $\sigma$ are restricted to be transverse, and we have used the combinations
 \begin{align}
\tilde{D}_T(z,k_T^2)&=\tilde{D'}_T(z,k_T^2)-{k_T^2\over 2M_\Lambda^2}\tilde{D}^\perp_T(z,k_T^2)\,,\\
\tilde{G}_T(z,k_T^2)&=\tilde{G'}_T(z,k_T^2)-{k_T^2\over 2M_\Lambda^2}\tilde{G}^\perp_T(z,k_T^2)\,,
\end{align}


Using the parameterizations of correlators in Eq.~(\ref{eq:hadtensor}) and the above identities, we can calculate the hadronic tensor in the process in which a lepton scatters off an unpolarized target producing a polarized $\Lambda$ hyperon. We obtain the complete results for the symmetric and antisymmetric part of the hadronic tensor:
\begin{align}\label{eq:W1}
2M W_S^{\mu\nu}=&2MW^{[MT]\mu\nu}_{S}+2z\int{d^2p_Td^2k_T}\delta^2(\bm p_T-\bm k_T +\bm q_T)\nonumber\displaybreak[0]\\
&\times\bigg{\{}-{(p_\perp^{\{\mu}\epsilon^{\nu\}\rho}_\perp S_{\Lambda\,\perp\rho}+S_{\Lambda\perp}^{\{\mu}\epsilon^{\nu\}\rho}_\perp p_{\perp\rho})\over 2M}h_1^\perp\,H_{1T}-{(p_\perp^{\{\mu}\epsilon^{\nu\}\rho}_\perp k_{\perp\rho}+k_{\perp}^{\{\mu}\epsilon^{\nu\}\rho}_\perp p_{\perp\rho})\over 2MM_\Lambda}h_1^\perp\,H_{1s}^\perp\nonumber\displaybreak[0]\\
&-(-g^{\mu\,\nu}_\perp (k_\perp\cdot p_\perp)+k^{\{\mu}_\perp p^{\nu\}}_\perp){1\over M\,M_\Lambda}h_1^\perp\,H_1^\perp+{t^{\{\mu}k_\perp^{\nu\}}\over Q}\left[{2xM\over M_\Lambda}hH_1^\perp\right]+{t^{\{\mu}p_\perp^{\nu\}}\over Q}\left[{M_\Lambda\over M}{2\tilde{H}\over z}\,h^\perp_1\right]\nonumber\displaybreak[0]\\
&+{t^{\{\mu}\,\epsilon^{\nu\}\rho}_\perp S_{\Lambda\perp\rho}\over Q}\left[-{p_\perp\cdot k_\perp\over M}2h_1^\perp\,{\tilde{H}^\perp_{T}\over z}+{M_\Lambda}f_1\,{2\tilde{D}_T\over z}-{2xM}{h\,H_{1T}}\right]\nonumber\displaybreak[0]\\
&+{t^{\{\mu}\,\epsilon^{\nu\}\rho}_\perp k_{\perp\rho}\over Q}\left[S_{\Lambda\,L}\,2f_1\,{\tilde{D}^\perp_{L}\over z}-{2(S_{\Lambda\perp}\cdot k_\perp)\over M_\Lambda}f_1\,{{D}^\perp_{T}\over z}-{M\over M_\Lambda}2xh\,{H^\perp_{1s}}+{(S_{\Lambda\,\perp}\cdot p_\perp)\over M}2h^\perp_1\,{\tilde{H}^\perp_{T}\over z}\right]\nonumber\displaybreak[0]\\
&+{t^{\{\mu}\,\epsilon^{\nu\}\rho}_\perp p_{\perp\rho}\over Q}\left[-2{M_\Lambda\over M}h_1^\perp{\tilde{H}_s\over z}-2xg^\perp\,G_{1s}\right]\bigg{\}}\displaybreak[0]\,,
\end{align}
and
\begin{align}\label{eq:W2}
2M W_A^{\mu\nu}=&2MW^{[MT]\mu\nu}_{A}+2z\int{d^2p_T\,d^2k_T}\delta^2(\bm p_T +\bm q_T-\bm k_T)\nonumber\displaybreak[0]\\
&\times\bigg{\{}i{t^{[\mu}k_\perp^{\nu]}\over Q}\left[-2f_1{\tilde{G}^\perp\over z}\right]+i{t^{[\mu}p_\perp^{\nu]}\over Q}\left[2{M_\Lambda\over zM}\tilde{E}h^\perp_1+2xg^\perp D_1\right]\nonumber\displaybreak[0]\\
&+i{t^{[\mu}\,\epsilon^{\nu]\rho}_\perp S_{\Lambda\,\perp\rho}\over Q}\left[{p_\perp\cdot k_\perp\over M_\Lambda}2xg^\perp D^\perp_{1T}+{p_\perp\cdot k_\perp\over M}2{h_1^\perp\,{\tilde{E}^\perp_{T}\over z}}+{M_\Lambda}2{f_1\,{\tilde{G}_T\over z}}\right]\nonumber\displaybreak[0]\\
&+i{t^{[\mu}\,\epsilon^{\nu]\rho}_\perp k_{\perp\rho}\over Q}\left[-{(S_{\Lambda\,\perp}\cdot p_\perp)\over M_\Lambda}2xg^\perp D^\perp_{1T}-{(S_{\Lambda\,\perp}\cdot p_\perp)\over M}2h^\perp_1\,{\tilde{E}^\perp_{T}\over z}\right]\nonumber\displaybreak[0]\\
&+i{t^{[\mu}\,\epsilon^{\nu]\rho}_\perp p_{\perp\rho}\over Q}\left[-2{S_{\Lambda\,L}M_\Lambda\,\over M}h_1^\perp{\tilde{E}_s\over z}\right]\bigg{\}}\displaybreak[0]\,,
\end{align}
where $2MW^{[MT]\mu\nu}_{S/A}$ denotes to the hadronic tensor for unpolarized $\Lambda$ production.
We find that $2MW^{[MT]\mu\nu}_{S/A}$ agree exactly with the results in Ref.~\cite{Mulders:1995dh}.
The rest part is our new results for polarized $\Lambda$ production.
We would like to point out that our results also satisfy the electromagnetic gauge invariance ($q_\mu\,W^{\mu\nu}=0$).
\section{The result of structure functions}

To find the number of independent structure functions in the leptoproduction of $\Lambda$ hyperon off an unpolarized target nucleon, we adopt a general analysis used in Refs.~\cite{Kotzinian:1994dv,Gliske:2014wba}.
The hadronic tensor $W_{\mu\nu}$ can be decomposed into scalar structure functions based on the polarization vectors of the virtual photon ($\epsilon^\mu$) and $\Lambda$ hyperon ($e^\mu$):
\begin{align}
& W^{(0)}_{\mu\nu}=\epsilon^a_\mu\epsilon^b_\nu\,H_{ab}^{(0)}\,,~W^{(S)}_{\mu\nu\rho}=\epsilon^a_\mu\epsilon^b_\nu e^i_\rho\,H_{abi}^{(S)}\,,
\end{align}
where $W^{(0)}_{\mu\nu}$ is the $\Lambda$-spin-independent part and $W^{(S)}_{\mu\nu}= S_\Lambda^\rho\,W^{(S)}_{\mu\nu\rho}$ is the $\Lambda$-spin-dependent part.
On the other hand, the scalar structure functions can be expressed as
\begin{align}
& H^{(0)}_{ab}=\epsilon^\mu_a\epsilon^\nu_b\,W_{\mu\nu}^{(0)}\,,~ H^{(S)}_{abc}=\epsilon^\mu_a\epsilon^\nu_b e^\rho_c\,W_{\mu\nu\rho}^{(S)}\,.
\end{align}

To simplify the calculation, here we adopt a special reference frame in which the $\Lambda$ hyperon is at rest and the momentum of the photon $\bm q$ is along the $z$ axis.
Thus the polarization vectors $\epsilon^\mu$ and $e^\mu$ can be chosen as
\begin{align}
&\epsilon_0^\mu={1\over Q}(q^3,0,0,q^0), &e_1^\mu=(1,0,0,0)\,,\notag\\
&\epsilon_1^\mu={1\over Q}(0,1,0,0), &e_1^\mu=(0,1,0,0)\,,\notag\\
&\epsilon_2^\mu={1\over Q}(0,0,1,0), &e_2^\mu=(0,0,1,0)\,,\\
&\epsilon_3^\mu={q^\mu\over Q}, &e_3^\mu=(0,0,0,1)\,.\notag
\end{align}

If there is no any restriction, in principle there are totally 160 real structure functions.
From the restrictions of the hermiticity, parity invariance\footnote{As is well-known, the decay of the $\Lambda$ hyperon to $p \pi$ violates parity invariance. However, in this work we only consider the case that the measured final state is the $\Lambda$ hyperon and we refrain from the complexity of Lambda decay.}, current conservation and the constraint $P_\Lambda \cdot S_\Lambda =0$, one has:
\begin{align}
 H_{ab}^{(0)}&=H_{ba}^{(0)\ast},~H_{abc}^{(S)}=H_{bac}^{(S)\ast}\,;\nonumber\\
 H_{ab}^{(0)}&=0, ~\textrm{if}~ a=2~ \textrm{and} ~b\neq2,\;~\textrm{or}~\,b=2  ~\textrm{and} ~a\neq2\,;\nonumber\\
 H_{abc}^{(S)}&=0, ~\textrm{if it contains an even number of indices 2}\,;\nonumber\\
 H_{ab0}^{(S)}&=H_{3bc}^{(S)}=H_{a3c}^{(S)}=H_{3b}^{(0)}=H_{a3}^{(0)}=0;\nonumber
\end{align}
These limits the structure functions to 18 independent real ones: five spin-independent ones $H_{00}^{(0)}$, $H_{11}^{(0)}$, $H_{22}^{(0)}$, $\Re H_{01}^{(0)}$, $\Im H_{10}^{(0)}$ and thirteen $\Lambda$-spin-dependent ones $H_{002}^{(S)}$, $\Re H_{012}^{(S)}$, $\Im H_{012}^{(S)}$, $\Re H_{021}^{(S)}$, $\Im H_{021}^{(S)}$, $\Re H_{023}^{(S)}$, $\Im H_{023}^{(S)}$, $\Re H_{121}^{(S)}$, $\Im H_{121}^{(S)}$, $H_{112}^{(S)}$, $\Re H_{123}^{(S)}$, $\Im H_{123}^{(S)}$, $H_{222}^{(S)}$, with $\Im$ and $\Re$ denoting the imaginary and real parts of the structure functions.

On the other hand, in the parton model the cross section of the process $\ell\, p\rightarrow \ell^\prime\, \Lambda\, X $ can be directly calculated from the contraction of the leptonic tensor (\ref{lepton}) with the hadronic tensor (\ref{eq:W1}) and (\ref{eq:W2}).
We find that up to twist-3, among 13 $\Lambda$-spin-dependent structure functions,  there are twelve non-zero ones that can be expressed as the convolution of the distribution functions and fragmentation functions.
Here we present the general expression of the cross section (for polarized $\Lambda$ production) in the $\gamma^* N$ collinear frame:
\begin{align}\label{cross}
{{d\sigma}\over{dx dy dz d\phi d\psi dP^2_{\Lambda\perp}}} &= {\alpha^2\over xyQ^2} \bigg{\{}S_{\Lambda_{||}}[(2-y)\sqrt{1-y}\sin{\phi_\Lambda} F_{UUL}^{\sin\phi_\Lambda}+(1-y)\sin{2\phi_\Lambda}F_{UUL}^{\sin2\phi_\Lambda}]\nonumber\\
&+S_{\Lambda_{||}}\lambda[y(1-{1\over 2}y)F_{LUL}+y\sqrt{1-y}\cos\phi_\Lambda\,F_{LUL}^{\cos\phi_\Lambda}]\nonumber\\
&+\left|S_{\Lambda_\perp}\right|[(1-y+{1\over 2}y^2)\sin(\phi_\Lambda-\phi_{S_{\Lambda}})F_{UUT}^{\sin(\phi_\Lambda-\phi_{S_\Lambda})}+(1-y)\sin(\phi_\Lambda+\phi_{S_{\Lambda}})
F_{UUT}^{\sin\phi_\Lambda+\phi_{S_{\Lambda}}}\nonumber\\
&+(1-y)\sin(3\phi_\Lambda-\phi_{S_{\Lambda}})F_{UUT}^{\sin(3\phi_\Lambda-\phi_{S_{\Lambda}})}
+(2-y)\sqrt{1-y}\sin(\phi_{S_{\Lambda}})F_{UUT}^{\sin\phi_{S_{\Lambda}}}\nonumber\\
&+(2-y)\sqrt{1-y}\sin(2\phi_\Lambda-\phi_{S_{\Lambda}})F_{UUT}^{\sin(2\phi_\Lambda-\phi_{S_{\Lambda}})}]\nonumber\\
&+\left|S_{\Lambda_\perp}\right|\lambda[y(1-{1\over 2}y)\cos(\phi_\Lambda-\phi_{S_{\Lambda}})F_{LUT}^{\cos(\phi_\Lambda-\phi_{S_{\Lambda}})}+y\sqrt{(1-y)}\cos(\phi_{S_{\Lambda}})F_{LUT}^{\cos\phi_{S_{\Lambda}}}\nonumber\\
&+y\sqrt{(1-y)}\cos(2\phi_\Lambda-\phi_{S_{\Lambda}})
F_{LUT}^{\cos(2\phi_\Lambda-\phi_{S_{\Lambda}})}\bigg{\}}\,,
\end{align}
where $F_{ABC}=F_{ABC}(x,z,P^2_{h\perp})$ denotes the structure functions with different angular modulations.
The subscripts $A, B$ and $C$ indicate the polarizations of the incoming lepton, the target nucleon and the produced $\Lambda$ hyperon, respectively; and we use $U$, $L$ and $T$ to denote unpolarized, longitudinally and transversely polarized particles.

The expressions of the structure functions in Eq.~(\ref{cross}) have the following explicit forms
\begin{align}
\begin{split}\label{eqstructure}
F^{\sin(\phi_{\Lambda}-\phi_{S_\Lambda})}_{UUT}&=\mathcal{I}\big{[}{\hat{\bm{h}}\cdot\bm{k}_{T}\over M_\Lambda}f_1D^\perp_{1T}\big{]}\,,
\end{split} \displaybreak[0]\\
\begin{split}
F^{\sin{2\phi_{\Lambda}}}_{UUL}&=\mathcal{I}\big{[}-{2(\hat{\bm{h}}\cdot\bm{p}_T)(\hat{\bm{h}}\cdot\bm{k}_T)-\bm{p}_T\cdot\bm{k}_T\over MM_\Lambda}h_1^\perp\,H_{1L}^\perp\big{]}\,,
\end{split} \displaybreak[0]\\
\begin{split}
F^{\sin{(\phi_{\Lambda}+\phi_{S_\Lambda})}}_{UUT}&=\mathcal{I}\big{[}-{\hat{\bm{h}}\cdot\bm{p}_T\over M}h_1^\perp\,H_{1}\big{]}\,,
\end{split} \displaybreak[0]\\
\begin{split}
F^{\sin{(3\phi_{\Lambda}-\phi_{S_\Lambda})}}_{UUT}&=\mathcal{I}\big{[}-{4(\hat{\bm{h}}{\cdot\bm{p}_T})(\hat{\bm{h}}{\cdot\bm{k}_T})^2
-\bm{k}^2_\perp(\hat{\bm{h}}{\cdot\bm{p}_T})-2(\hat{\bm{h}}{\cdot\bm{k}_T})(\bm{k}_T{\cdot\bm{p}_T})\over 2MM^2_\Lambda}h_1^\perp\,H_{1T}^\perp\big{]}\,,
\end{split} \displaybreak[0]\\
\begin{split}
F_{LUL}&=\mathcal{I}\big{[}f_1G_{1L}\big{]}\,,
\end{split} \displaybreak[0]\\
\begin{split}
F^{\cos{(\phi_\Lambda-\phi_{S_\Lambda})}}_{LUT}&=\mathcal{I}\big{[}{\hat{\bm{h}}\cdot\bm{k}_T\over M_\Lambda}f_1G_{1T}\big{]}\,,
\end{split} \displaybreak[0]\\
\begin{split}
F^{\sin{\phi_{\Lambda}}}_{UUL}&={2M\over Q}\mathcal{I}\bigg{\{}{\hat{\bm{h}}\cdot \bm{k}_T\over M_\Lambda}({M_\Lambda\over M}f_1\,{\tilde{D}^\perp_L\over z}-xh\,H^\perp_{1L})-{\hat{\bm{h}}\cdot \bm{p}_T\over M}({M_\Lambda\over M}h^\perp_1{\tilde{H}_L}+xg^\perp\,G_{1L})\bigg{\}}\,,
\end{split} \displaybreak[0]\\
\begin{split}
F^{\sin{(2\phi_\Lambda-\phi_{S_\Lambda})}}_{UUT}&={2M\over Q}\mathcal{I}\bigg{\{}{2(\hat{\bm{h}}\cdot \bm{k}_T)^2-\bm{k}^2_T\over 2M^2_\Lambda}({M_\Lambda\over M}f_1\,{\tilde{D}^\perp_{T}\over z}-xh\,H^\perp_{1T})\\
&+{2(\hat{\bm{h}}\cdot \bm{k}_T)(\hat{\bm{h}}\cdot \bm{p}_T)-\bm{k}_T\cdot\bm{p}_T \over 2M\,M_\Lambda}\big{[}(xf^\perp\,D^\perp_{1T}-{M_\Lambda\over M}h_1^\perp\,{\tilde{H}^\perp_T\over z})-({M_\Lambda\over M}h_1^\perp{\tilde{H}_{T}\over z}+xg^\perp\,G_{1T})\big{]}\bigg{\}}\,,
\end{split} \displaybreak[0]\\
\begin{split}
F^{\sin{\phi_{S_\Lambda}}}_{UUT}&={2M\over Q}\mathcal{I}\bigg{\{}({M_\Lambda\over M}f_1\,{\tilde{D}_T\over z}-x{h\,H_{1}})+{\bm{k}_T\,\cdot\bm{p}_T \over 2MM_\Lambda}[({M_\Lambda\over M}h_1^\perp\tilde{H}_T^\perp-xf_1\,D^\perp_{1T})-({M_\Lambda\over M}h_1^\perp{\tilde{H}_{T}\over z}-xg^\perp\,G_{1T})]\bigg{\}}\,,
\end{split}\displaybreak[0]\\
\begin{split}
F^{\cos{\phi_{S_\Lambda}}}_{LUT}&={2M\over Q}\mathcal{I}\bigg{\{}-({M_\Lambda\over M}f_1\,{\tilde{G}_T\over z}+xe\,H_{1})+{\bm{p}_T\cdot \bm{k}_T\over 2MM_\Lambda}[({M_\Lambda\over M}h^\perp_1\,{\tilde{E}^\perp_{T}\over z}+xg^\perp D^\perp_{1T})+({M_\Lambda\over M}h_1^\perp{\tilde{E}_{T}\over z}-xf^\perp\, G_{1T})]\bigg{\}}\,,
\end{split} \displaybreak[0]\\
\begin{split}
F^{\cos{(2\phi_\Lambda-\phi_{S_\Lambda})}}_{LUT}&={2M \over Q}\mathcal{I}\bigg{\{}-{2(\hat{\bm{h}}\cdot \bm{k}_T)^2-\bm{k}_T^2\over 2M^2_\Lambda}({M_\Lambda\over M}f_1\,{\tilde{G}^\perp_{T}\over z}+xe\,H^\perp_{1T})\\
&-{2(\hat{\bm{h}}\cdot \bm{k}_T)(\hat{\bm{h}}\cdot \bm{p}_T)-\bm{k}_T\cdot \bm{p}_T\over 2M\,M_\Lambda}[(xg^\perp\,D^\perp_{1T}+{M_\Lambda\over M}h^\perp_1\,{\tilde{E}^\perp_{T}\over z})-({M_\Lambda\over M}h_1^\perp{\tilde{E}_{T}\over z}-xf^\perp\,G_{1T})]\bigg{\}}\,,
\end{split} \displaybreak[0]\\
\begin{split}\label{eqstru}
F^{\cos{\phi_\Lambda}}_{LUL}&={2M \over Q}\mathcal{I}\bigg{\{}{{\hat{\bm{h}}\cdot \bm{p}_T}\over M}({M_\Lambda\over M}\,h_1^\perp{\tilde{E}_L\over z}-x\,f^\perp G_{1L})-{\hat{\bm{h}}\cdot \bm{k}_T\over M_\Lambda}({M_\Lambda\over M}f_1\,{\tilde{G}^\perp_L\over z}+\,xe\,H^\perp_{1L})\bigg{\}}\,,
\end{split}
\end{align}
In the expressions of the structure functions we introduce the normalized vector $\hat{\bm h}= \bm P_{h\perp}/|\bm P_{h\perp}|$ and the convolution integral
\begin{align}
\mathcal{I}\big{[}\omega\,f\,D\big{]}=x\sum_a\,e^2_a\,\int{d^2\bm{p}_T\,d^2\bm{k}_T}\delta^{(2)}({\bm{p}_T-\bm{k}_T-
\bm{P}_{\Lambda\perp}/z})\,\omega(\bm{p}_T,\bm{k}_T)
f^a(x,p^2_T)D^a(z,k^2_T)\,.
\end{align}
Here, $\omega(\bm p_T,\bm k_T)$ is an arbitrary function and the summation runs over quarks and anti-quarks.
In above results, we have applied the equation of motion relations~\cite{Mulders:1995dh} among the twist-2 and twist-3 TMD PDFs and FFs.
Thus we can find a feature that PDFs and FFs do not appear in a symmetric fashion: there are only twist-3 PDFs without tilde and twist-3 FFs with a tilde.
The reason of this asymmetry is that in Eq.~(\ref{cross}) the structure functions are introduced by using an asymmetric way in the $\gamma^*\,N$ collinear frames rather than the $\Lambda N$ collinear frames.

In the following, we will briefly discuss our results.

1. The contributions of the new polarized FFs $\tilde D^\perp_T$, $\tilde E^\perp_T$ have to be taken into account in the calculation of structure functions.
Our calculation shows that $\tilde D^\perp_T$ appear in the $\phi_{S_\Lambda}$-dependent structure functions $F_{UUT}^{\sin(2\phi_\Lambda-\phi_{S_\Lambda})}$, while  $\tilde E^\perp_T$ appears in $F_{LUT}^{\cos\phi_{S_\Lambda}}$ and $F_{LUT}^{\cos(2\phi_\Lambda-\phi_{S_\Lambda})}$.
The twist-3 TMDs $g^\perp$ gives contributions to all the twist-3 structure functions except $F_{LUL}^{\cos\phi_{S_\Lambda}}$.

2. The leading twist structure functions in our results are corresponding to Table. \uppercase\expandafter{\romannumeral3} of Ref.~\cite{Boer:1997nt} and Eq.(38) in Ref.~\cite{Boer:1999uu}.
The six twist-3 structure functions in our results show some similarity compared with the expressions in Ref.~\cite{Bacchetta:2006tn}, while the role of PDFs and FFs have being reversed in the former.

3. Our results show that the polarized FFs play important roles in the quark fragmenting to $\Lambda$ hyperon, i.e., the T-odd FF $D^\perp_{1T}$ have been studied in Refs.~\cite{Anselmino:2001js,Boer:2010ya} and our results show that it can be measured in $\sin(\phi_{\Lambda}-\phi_{S_\Lambda})$ asymmetry.
Because the TMD FFs are found to be universal in different process~\cite{Meissner:2008yf,Collins:2004nx}, the FFs can be also applied to study the $\Lambda$ production in $e^+\,e^-$ annihilation process.

4. If one neglects the quark-gluon-quark functions (the functions with a tilde) and T-odd distribution functions $h_1^\perp$, $g^\perp$, $h$, then five structure functions at twist-3 have the approximated forms:
\begin{align}
 \begin{split}
F^{\sin{(2\phi_\Lambda-\phi_{S_\Lambda})}}_{UUT}&\approx {2M\over Q}\mathcal{I}\bigg{\{}{2(\hat{\bm{h}}\cdot \bm{k}_T)(\hat{\bm{h}}\cdot \bm{p}_T)-\bm{k}_T\cdot\bm{p}_T \over 2M\,M_\Lambda}f_1\,D^\perp_{1T}\bigg{\}}\,,
\end{split} \displaybreak[0]\\
\begin{split}
F^{\sin{\phi_{S_\Lambda}}}_{UUT}&\approx {2M\over Q}\mathcal{I}\bigg{\{}-{\bm{k}_T\,\cdot\bm{p}_T \over 2MM_\Lambda}xf_1\,D^\perp_{1T}\bigg{\}}\,,
\end{split}\displaybreak[0]\\
\begin{split}
F^{\cos{\phi_{S_\Lambda}}}_{LUT}&\approx {2M\over Q}\mathcal{I}\bigg{\{}-{m\over M}f_1\,H_{1}-{\bm{p}_T\cdot \bm{k}_T\over 2MM_\Lambda}f_1\, G_{1T}\bigg{\}}\,,
\end{split} \displaybreak[0]\\
\begin{split}
F^{\cos{(2\phi_\Lambda-\phi_{S_\Lambda})}}_{LUT}&\approx {2M \over Q}\mathcal{I}\bigg{\{}-{2(\hat{\bm{h}}\cdot \bm{k}_T)^2-\bm{k}_T^2\over 2M^2_\Lambda}{m\over M}f_1\,H^\perp_{1T}-{2(\hat{\bm{h}}\cdot \bm{k}_T)(\hat{\bm{h}}\cdot \bm{p}_T)-\bm{k}_T\cdot \bm{p}_T\over 2M\,M_\Lambda}f_1\,G_{1T}\bigg{\}}\,,
\end{split} \displaybreak[0]\\
\begin{split}
F^{\cos{\phi_\Lambda}}_{LUL}&\approx {2M \over Q}\mathcal{I}\bigg{\{}-{{\hat{\bm{h}}\cdot \bm{p}_T}\over M}f_1\,G_{1L}-{\hat{\bm{h}}\cdot \bm{k}_T\over M_\Lambda}{m\over M}f_1\,H^\perp_{1L}\bigg{\}}\,.
\end{split}
\end{align}
Thus, these results are similar to the Cahn effect in the unpolarized SIDIS.
They occur when the intrinsic transverse momentum is included in distribution and fragmentation functions.

At last, we perform the integration over the transverse momentum $P_{\Lambda\perp}$ for the structure functions given above, the non-vanished integrated structure functions are as follows~\cite{Boer:1997nt}
\begin{align}
\begin{split}
F_{LUL}(x,z)&=x\sum_a\,e_a^2\,f_1(x)G_{1}(z)\,,
\end{split} \displaybreak[0]\\
\begin{split}\label{eq:cf}
F^{\sin{\phi_{S_\Lambda}}}_{UUT}(x,z)&=x\sum_a\,e_a^2{2M_\Lambda\over Q}f_1(x)\,{\tilde{D}_T(z)\over z}\,,
\end{split}\displaybreak[0]\\
\begin{split}
F^{\cos{\phi_{S_\Lambda}}}_{LUT}(x,z)&=-x\sum_a\,e_a^2{2M\over Q}\bigg({M_\Lambda\over M}f_1(x)\,{\tilde{G}_T(z)\over z}+xe(x)\,H_{1}(z)\bigg)\,,\label{eq:eh}
\end{split}
\end{align}
where the functions on r.h.s are given by
\begin{align}
f_1(x)=\int{d^2\bm{p}_T f_1(x,p^2_T)}, ~~~~ D_1(z)=z^2\int{d^2\bm{k}_T D_1(z,k^2_T)}\,,
\end{align}
and so forth for the other functions. In Eq.~(\ref{eq:cf}), there is no contribution from the distribution $h$, since T-odd PDFs vanish under time reversal:
\begin{align}
\int{d^2\bm{p}_T h(x,p^2_T)}=0\,.\label{int}
\end{align}
The contribution from $\tilde{D}_T(z)$ still remains due to the final state interaction effects~\cite{Barone:2001sp} during fragmentation.
Thus, the measurement of the structure function $F^{\sin{\phi_{S_\Lambda}}}_{UUT}(x,z)$ provides a unique opportunity to explore the T-odd FF $\tilde{D_T}(z)$.
Furthermore, Eq.~(\ref{eq:eh}) shows a
possible way to observe the function $e(x)$ in the leptoproduction of transversely polarized $\Lambda$, provided the twist-3 fragmentation function $\tilde{G}_T(z)$ is negligible.
It is also worth pointing out that the structure function $F_{LUL}$ is associated with the longitudinal spin transfer from longitudinally polarized lepton beam to the $\Lambda$ hyperon through the convolution of $f_1(x)$ and $G_1(z)$~\cite{Jaffe:1996wp,Ma:2000uv,Mulders:1995dh}.

Finally, a general remark on $\Lambda$ fragmentation functions is in order.
In the entire paper, we do not consider the decay of the $\Lambda$ hyperon to $p \pi$.
In reality, $\Lambda$ fragmentation observables should be contained in the more general dihadron fragmentation observables, because in the end the $\Lambda$ hyperon decays to two hadrons, and this decay acts as a polarization analyzer.
Specifically, the polarization information of the $\Lambda$ hyperon during fragmentation might be explored by the correlation $(\bm P_p \wedge \bm P_\pi)\cdot \bm S$, where $\bm P_p$ and $\bm P_\pi$ are the momenta of the decayed proton and pion, and $\bm S$ is the spin vector of the $\Lambda$ hyperon.
This is similar to the correlation $(\bm P_1 \wedge \bm P_2)\cdot \bm s_q $ associated with the quark dihadron fragmentation function $q\rightarrow h_1 h_2$ which was proposed to access the transversity distribution of the nucleon.
Our calculation may be extended to the case of dihadron production in SIDIS, and we reserve it as a future study.

\section{conclusion}

SIDIS has been recognized as an very useful tool to study the
spin structure of $\Lambda$ hyperon.
In this work, we have studied the production of polarized $\Lambda$ hyperon by unpolarized or longitudinally lepton beam scattered off an unpolarized nucleon target.
Using the complete decomposition of the parton correlation functions for fragmentation up to twist three,
we have presented the tree-level result of the cross section for the process $\ell + N \rightarrow \ell^\prime + \Lambda +X$ at order $1/Q$, based on the TMD framework.
We find that, among a total of twelve non-zero polarized structure functions, seven of them are at twist-three level and can be expressed as convolutions of twist-two and twist-three TMD PDFs and FFs.
We give the complete expressions for these structure functions, each of which there are
several twist-three functions that contribute.
In our analysis we also include the T-odd TMD PDFs $g^\perp$ and FFs $\tilde{E}^\perp_T$ and $\tilde{D}^\perp_T$, which were not taken into account in previous studies and may contribute in spin asymmetries in polarized $\Lambda$ production.
The measurements of $\ell + N \rightarrow \ell^\prime + \Lambda +X$ thus
can provide useful observables to understand the fragmentation mechanism and polarization phenomena of the $\Lambda$ hyperon.

\section{Acknowledgements}
This work is partially supported by the National Natural Science
Foundation of China (Grants No.~11575043 and No.~11120101004), by the Fundamental Research Funds for the Central Universities, and by the Qing Lan Project.


\begin{thebibliography}{99}

 \bibitem{Bunce:1976yb}
  G.~Bunce {\it et al.},
  Phys.\ Rev.\ Lett.\  {\bf 36}, 1113 (1976).

\bibitem{Burkardt:1993zh}
  M.~Burkardt and R.~L.~Jaffe,
  Phys.\ Rev.\ Lett.\  {\bf 70}, 2537 (1993)
  [hep-ph/9302232].

\bibitem{Jaffe:1996wp}
  R.~L.~Jaffe,
  Phys.\ Rev.\ D {\bf 54}, no. 11, R6581 (1996)
  [hep-ph/9605456].

\bibitem{Ma:2000uv}
  B.~Q.~Ma, I.~Schmidt, J.~Soffer and J.~J.~Yang,
  Phys.\ Lett.\ B {\bf 488}, 254 (2000)
  [hep-ph/0005210].

  \bibitem{Anselmino:2001ps}
  M.~Anselmino, M.~Boglione, U.~D'Alesio, E.~Leader and F.~Murgia,
  Phys.\ Lett.\ B {\bf 509}, 246 (2001)
  [hep-ph/0102119].

\bibitem{Zhou:2009mx}
  S.~s.~Zhou, Y.~Chen, Z.~t.~Liang and Q.~h.~Xu,
  Phys.\ Rev.\ D {\bf 79}, 094018 (2009)
  [arXiv:0902.1883 [hep-ph]].

 \bibitem{Airapetian:1999sh}
  A.~Airapetian {\it et al.} [HERMES Collaboration],
  Phys.\ Rev.\ D {\bf 64} (2001) 112005
  [hep-ex/9911017].


\bibitem{Belostotsky:1999kj}
  S.~L.~Belostotsky {\it et al.} [HERMES Collaboration],
  Czech.\ J.\ Phys.\  {\bf 50S1}, 45 (2000).

  \bibitem{Bernreuther:2000wf}
  S.~Bernreuther [HERMES Collaboration],
  AIP Conf.\ Proc.\  {\bf 570}, 504 (2001).

\bibitem{Airapetian:2006ee}
  A.~Airapetian {\it et al.} [HERMES Collaboration],
  Phys.\ Rev.\ D {\bf 74}, 072004 (2006)
  [hep-ex/0607004].


  \bibitem{Naryshkin:2007zza}
  Y.~Naryshkin [HERMES Collaboration],
  Nucl.\ Phys.\ Proc.\ Suppl.\  {\bf 167}, 62 (2007).

\bibitem{Belostotski:2011zzb}
  S.~Belostotski {\it et al.} [HERMES Collaboration],
  Nucl.\ Phys.\ Proc.\ Suppl.\  {\bf 210-211} (2011) 111.









\bibitem{Sapozhnikov:2005sc}
  M.~G.~Sapozhnikov [COMPASS Collaboration],
  hep-ex/0503009.

 \bibitem{Alexakhin:2005dz}
  V.~Y.~Alexakhin [COMPASS Collaboration],
  hep-ex/0502014.


\bibitem{Alekseev:2009ab}
  M.~Alekseev {\it et al.} [COMPASS Collaboration],
  Eur.\ Phys.\ J.\ C {\bf 64}, 171 (2009)
  [arXiv:0907.0388 [hep-ex]].

  \bibitem{Anselmino:2001js}
  M.~Anselmino, D.~Boer, U.~D'Alesio and F.~Murgia,
  Phys.\ Rev.\ D {\bf 65}, 114014 (2002)
  [hep-ph/0109186].

\bibitem{Boer:2010ya}
  D.~Boer, Z.~B.~Kang, W.~Vogelsang and F.~Yuan,
  Phys.\ Rev.\ Lett.\  {\bf 105}, 202001 (2010)
  [arXiv:1008.3543 [hep-ph]].

\bibitem{Kanazawa:2015jxa}
  K.~Kanazawa, A.~Metz, D.~Pitonyak and M.~Schlegel,
  Phys.\ Lett.\ B {\bf 744}, 385 (2015)
  [arXiv:1503.02003 [hep-ph]].

  \bibitem{Bellwied:2002rg}
  R.~Bellwied [E896 Collaboration],
  Nucl.\ Phys.\ A {\bf 698} (2002) 499.

  \bibitem{Adler:2002pb}
  C.~Adler {\it et al.} [STAR Collaboration],
  Phys.\ Rev.\ Lett.\  {\bf 89} (2002) 132301
  [hep-ex/0205072].

  \bibitem{Abelev:2006cs}
  B.~I.~Abelev {\it et al.} [STAR Collaboration],
  Phys.\ Rev.\ C {\bf 75} (2007) 064901
  [nucl-ex/0607033].

  \bibitem{ATLAS:2014ona}
  G.~Aad {\it et al.} [ATLAS Collaboration],
  Phys.\ Rev.\ D {\bf 91} (2015) no.3,  032004
  [arXiv:1412.1692 [hep-ex]].


  \bibitem{Rith:2007zz}
  K.~Rith [HERMES Collaboration],
  AIP Conf.\ Proc.\  {\bf 915}, 445 (2007).

  \bibitem{Ferrero:2007zz}
  A.~Ferrero,
  AIP Conf.\ Proc.\  {\bf 915}, 436 (2007).

\bibitem{Airapetian:2014tyc}
  A.~Airapetian {\it et al.} [HERMES Collaboration],
  Phys.\ Rev.\ D {\bf 90}, no. 7, 072007 (2014)
  [arXiv:1406.3236 [hep-ex]].


\bibitem{Karyan:2016rls}
  G.~Karyan [HERMES Collaboration],
  Int.\ J.\ Mod.\ Phys.\ Conf.\ Ser.\  {\bf 40}, 1660067 (2016).

  \bibitem{Goeke:2003az}
  K.~Goeke, A.~Metz, P.~V.~Pobylitsa and M.~V.~Polyakov,
  Phys.\ Lett.\ B {\bf 567}, 27 (2003)
  [hep-ph/0302028].

  \bibitem{Bacchetta:2004zf}
  A.~Bacchetta, P.~J.~Mulders and F.~Pijlman,
  Phys.\ Lett.\ B {\bf 595}, 309 (2004)
  [hep-ph/0405154].

  \bibitem{Goeke:2005hb}
  K.~Goeke, A.~Metz and M.~Schlegel,
  Phys.\ Lett.\ B {\bf 618}, 90 (2005)
  [hep-ph/0504130].

  \bibitem{Mulders:1995dh}
  P.~J.~Mulders and R.~D.~Tangerman,
  Nucl.\ Phys.\ B {\bf 461}, 197 (1996)
  Erratum: [Nucl.\ Phys.\ B {\bf 484}, 538 (1997)]
  [hep-ph/9510301].

  \bibitem{Boer:1997nt}
  D.~Boer and P.~J.~Mulders,
  Phys.\ Rev.\ D {\bf 57}, 5780 (1998)
  [hep-ph/9711485].

  \bibitem{Bacchetta:2004jz}
  A.~Bacchetta, U.~D'Alesio, M.~Diehl and C.~A.~Miller,
  Phys.\ Rev.\ D {\bf 70}, 117504 (2004)
  [hep-ph/0410050].

  \bibitem{Bacchetta:2006tn}
  A.~Bacchetta, M.~Diehl, K.~Goeke, A.~Metz, P.~J.~Mulders and M.~Schlegel,
  JHEP {\bf 0702}, 093 (2007)
  [hep-ph/0611265].

  \bibitem{Boer:1999uu}
  D.~Boer, R.~Jakob and P.~J.~Mulders,
  Nucl.\ Phys.\ B {\bf 564}, 471 (2000)
  [hep-ph/9907504].


\bibitem{Gliske:2014wba}
  S.~Gliske, A.~Bacchetta and M.~Radici,
  Phys.\ Rev.\ D {\bf 90}, 114027 (2014)
  Erratum: [Phys.\ Rev.\ D {\bf 91}, 019902 (2015)]
  [arXiv:1408.5721 [hep-ph]].

 \bibitem{Diehl:2005pc}
  M.~Diehl and S.~Sapeta,
  Eur.\ Phys.\ J.\ C {\bf 41}, 515 (2005)
  [hep-ph/0503023].

  \bibitem{Collins:2004nx}
  J.~C.~Collins and A.~Metz,
  Phys.\ Rev.\ Lett.\  {\bf 93}, 252001 (2004)
  [hep-ph/0408249].

  \bibitem{Ji:2004xq}
  X.~d.~Ji, J.~P.~Ma and F.~Yuan,
  Phys.\ Lett.\ B {\bf 597}, 299 (2004)
  [hep-ph/0405085].

  \bibitem{Ji:2004wu}
  X.~d.~Ji, J.~p.~Ma and F.~Yuan,
  Phys.\ Rev.\ D {\bf 71}, 034005 (2005)
  [hep-ph/0404183].

 \bibitem{Aybat:2011zv}
  S.~M.~Aybat and T.~C.~Rogers,
  Phys.\ Rev.\ D {\bf 83}, 114042 (2011)
  [arXiv:1101.5057 [hep-ph]].

  \bibitem{Curci:1980uw}
  G.~Curci, W.~Furmanski and R.~Petronzio,
  Nucl.\ Phys.\ B {\bf 175}, 27 (1980).

  \bibitem{Collins:1997sr}
  J.~C.~Collins,
  Phys.\ Rev.\ D {\bf 57}, 3051 (1998)
  Erratum: [Phys.\ Rev.\ D {\bf 61}, 019902 (2000)]
  [hep-ph/9709499].

 \bibitem{Collins:1989gx}
  J.~C.~Collins, D.~E.~Soper and G.~F.~Sterman,
  Adv.\ Ser.\ Direct.\ High Energy Phys.\  {\bf 5}, 1 (1989)
  [hep-ph/0409313].

 \bibitem{Collins:1985ue}
  J.~C.~Collins, D.~E.~Soper and G.~F.~Sterman,
  Nucl.\ Phys.\ B {\bf 261}, 104 (1985).

  \bibitem{Ma:2008cj}
  J.~P.~Ma and H.~Z.~Sang,
  Phys.\ Lett.\ B {\bf 676}, 74 (2009)
  [arXiv:0811.0224 [hep-ph]].

  \bibitem{Boer:2003cm}
  D.~Boer, P.~J.~Mulders and F.~Pijlman,
  Nucl.\ Phys.\ B {\bf 667}, 201 (2003)
  [hep-ph/0303034].

  \bibitem{Boer:1997mf}
  D.~Boer, R.~Jakob and P.~J.~Mulders,
  Nucl.\ Phys.\ B {\bf 504}, 345 (1997)
  [hep-ph/9702281].

  \bibitem{Metz:2016swz}
  A.~Metz and A.~Vossen,
  arXiv:1607.02521 [hep-ex].

  \bibitem{Belitsky:2002sm}
  A.~V.~Belitsky, X.~Ji and F.~Yuan,
  Nucl.\ Phys.\ B {\bf 656}, 165 (2003)
  [hep-ph/0208038].

  \bibitem{Ji:2002aa}
  X.~d.~Ji and F.~Yuan,
  Phys.\ Lett.\ B {\bf 543}, 66 (2002)
  [hep-ph/0206057].

  \bibitem{Collins:2002kn}
  J.~C.~Collins,
  Phys.\ Lett.\ B {\bf 536}, 43 (2002)
  [hep-ph/0204004].

  \bibitem{Bacchetta:2005rm}
  A.~Bacchetta, C.~J.~Bomhof, P.~J.~Mulders and F.~Pijlman,
  Phys.\ Rev.\ D {\bf 72}, 034030 (2005)
  [hep-ph/0505268].

  \bibitem{Bomhof:2004aw}
  C.~J.~Bomhof, P.~J.~Mulders and F.~Pijlman,
  Phys.\ Lett.\ B {\bf 596}, 277 (2004)
  [hep-ph/0406099].

  \bibitem{Bomhof:2006dp}
  C.~J.~Bomhof, P.~J.~Mulders and F.~Pijlman,
  Eur.\ Phys.\ J.\ C {\bf 47}, 147 (2006)
  [hep-ph/0601171].

\bibitem{Collins:2011zzd}
  J.~Collins,
  (Cambridge monographs on particle physics, nuclear physics and cosmology. 32)

\bibitem{Pitonyak:2013dsu}
  D.~Pitonyak, M.~Schlegel and A.~Metz,
  Phys.\ Rev.\ D {\bf 89}, 054032 (2014)
  [arXiv:1310.6240 [hep-ph]].

  \bibitem{Collins:1992kk}
  J.~C.~Collins,
  Nucl.\ Phys.\ B {\bf 396}, 161 (1993)
  [hep-ph/9208213].
  \bibitem{Bacchetta:2007wc}
  A.~Bacchetta, L.~P.~Gamberg, G.~R.~Goldstein and A.~Mukherjee,
  Phys.\ Lett.\ B {\bf 659}, 234 (2008)
  [arXiv:0707.3372 [hep-ph]].

  \bibitem{Yang:2016mxl}
  Y.~Yang, Z.~Lu and I.~Schmidt,
  Phys.\ Lett.\ B {\bf 761}, 333 (2016)
  [arXiv:1607.01638 [hep-ph]].

\bibitem{Lu:2015wja}
  Z.~Lu and I.~Schmidt,
  Phys.\ Lett.\ B {\bf 747}, 357 (2015)
  [arXiv:1501.04379 [hep-ph]].

  \bibitem{Metz:2012ct}
  A.~Metz and D.~Pitonyak,
  Phys.\ Lett.\ B {\bf 723}, 365 (2013)
  [arXiv:1212.5037 [hep-ph]].

\bibitem{Kotzinian:1994dv}
  A.~Kotzinian,
  Nucl.\ Phys.\ B {\bf 441}, 234 (1995)
  [hep-ph/9412283].

  \bibitem{Barone:2001sp}
  V.~Barone, A.~Drago and P.~G.~Ratcliffe,
  Phys.\ Rept.\  {\bf 359}, 1 (2002)
  [hep-ph/0104283].

\bibitem{Meissner:2008yf}
  S.~Meissner and A.~Metz,
  Phys.\ Rev.\ Lett.\  {\bf 102}, 172003 (2009)
  [arXiv:0812.3783 [hep-ph]].

\end{thebibliography}
\end{document}